\documentclass{aa}
\usepackage{graphicx}
\usepackage{natbib}
\usepackage{amssymb}
\begin{document}

\title{Mass loss rates of a sample of irregular and semiregular M-type
AGB-variables}


\author{H.~Olofsson\inst{1} \and D.~Gonz{\'a}lez Delgado\inst{1}
\and F.~Kerschbaum\inst{2} \and F.~L.~Sch\"oier\inst{3}}

\institute{Stockholm Observatory, SCFAB, SE-10691 Stockholm, Sweden
\and Institut f\"ur Astronomie, T\"urkenschanzstrasse 17, 1180 Wien, Austria
\and Leiden Observatory, PO Box 9513, 2300 RA Leiden, The Netherlands}

\offprints{H.~Olofsson (hans@astro.su.se)}
\date{A\&A accepted}
    
\abstract{We have determined mass loss rates and gas expansion velocities for a
sample of 69 M-type irregular (IRV; 22 objects) and semiregular (SRV; 47 objects)
AGB-variables using a radiative transfer code to model their circumstellar
CO radio line emission.  We believe that this sample is representative
for the mass losing stars of this type.  The (molecular hydrogen) mass loss rate
distribution has a median value of
2.0$\times$10$^{-7}$\,M$_{\odot}$\,yr$^{-1}$, and a minimum of
2.0$\times$10$^{-8}$\,M$_{\odot}$\,yr$^{-1}$ and a maximum of
8$\times$10$^{-7}$\,M$_{\odot}$\,yr$^{-1}$.  M-type IRVs and SRVs with
a mass loss rate in excess of
5$\times$10$^{-7}$\,M$_{\odot}$\,yr$^{-1}$ must be very rare, and
among these mass losing stars the number of sources with mass loss
rates below a few 10$^{-8}$\,M$_{\odot}$\,yr$^{-1}$ must be small.  We
find no significant difference between the IRVs and the SRVs in terms
of their mass loss characteristics.  Among the SRVs the mass loss rate
shows no dependence on the period.  Likewise the mass loss rate shows
no correlation with the stellar temperature.

The gas expansion velocity distribution has a median of
7.0\,km\,s$^{-1}$, and a minimum of 2.2\,km\,s$^{-1}$ and a maximum of
14.4\,km\,s$^{-1}$.  No doubt, these objects sample the low gas
expansion velocity end of AGB winds.  The fraction of objects with low
gas expansion velocities is very high, about 30\% have velocities
lower than 5\,km\,s$^{-1}$, and there are objects with velocities
lower than 3\,km\,s$^{-1}$: \object{V584~Aql}, \object{T~Ari},
\object{BI~Car}, \object{RX~Lac}, and \object{L$^2$~Pup}.  The mass
loss rate and the gas expansion velocity correlate well, a result in line with
theoretical predictions for an optically thin, dust-driven wind.

In general, the model produces line profiles which acceptably fit the
observed ones.  An exceptional case is \object{R~Dor}, where the high-quality,
observed line profiles are essentially flat-topped, while the model
ones are sharply double-peaked.

The sample contains four sources with distinctly double-component CO line
profiles, i.e., a narrow feature centered on a broader feature:
\object{EP~Aqr}, \object{RV~Boo}, \object{X~Her}, and \object{SV~Psc}. 
We have modelled the two components separately for each star and
derive mass loss rates and gas expansion velocities.

We have compared the results of this M-star sample with a similar
C-star sample analysed in the same way.  The mass loss rate
characteristics are very similar for the two samples.  On the contrary, the gas
expansion velocity distributions are clearly different.  In
particular, the number of low-velocity sources is much higher in the
M-star sample.  We found no example of the sharply double-peaked CO
line profile, which is evidence of a large, detached CO-shell, among
the M-stars.  About 10\% of the C-stars show this phenomenon.
    
}

\maketitle

\section{Introduction}

It has been firmly established that mass loss from the surface is a
very important process during the final stellar evolution of low- and
intermediate-mass stars, i.e., on the asymptotic giant branch (AGB). 
The mass loss seems to occur irrespective of the chemistry (C/O$<$1
or $>$1) or the variability pattern (irregular, semi-regular, or
regular) of the star.  Beyond these general conclusions the situation
becomes more uncertain \citep{olof99}.  Of importance for comparison
with mass loss models and for the understanding of AGB-stars in a
broader context (e.g., their contribution to the chemical evolution of
galaxies) is to establish the mass loss rate dependence on stellar
parameters, such as main sequence mass, luminosity, effective
temperature, pulsational pattern, metallicity, etc., and its evolution
with time for individual sources.  A crude picture has emerged where
the average mass loss rate increases as the star evolves along the
AGB, and where the final mass loss rate reached increases with the
main sequence mass.  In addition, there is evidence of time variable mass loss
\citep{haleetal97,maurhugg00}, and even highly episodic mass loss
\citep{olofetal00}.  There is a dependence on luminosity in the
expected way, i.e., an increase with increasing luminosity, but it is
uncertain how strong it is.  The same applies to the effective
temperature where a decrease with increasing temperature is expected. 
Regular pulsators clearly have higher mass loss rates than stars with
less regular pulsation patterns.  The dependence of the total mass
loss rate on metallicity appears to be weak, but the dust mass loss
rate decreases with decreasing metallicity.  See \citet{habi96} for a
summary of evidence in favour of this general outline.

The mechanism behind the mass loss remains unknown, even though there
are strong arguments in favour of a wind which is basically
pulsation-driven, and where the highest mass loss rates and gas
expansion velocities are reached through the addition of radiation
pressure on dust \citep{hoefdorf97,wintetal00b}.  A way to study this
problem is to use samples of low mass loss rate stars for which
stellar parameters can be reasonably estimated using traditional
methods.  These samples also contain objects with quite varying
pulsational characteristics, and has, as it turned out, quite varying
circumstellar characteristics.  \citet{olofetal93a} presented such a
study of low mass loss rate C-stars using CO multi-transition radio
data.  These data were subsequently analysed in more detail by
\citet{schoolof01} using a radiative transfer model.  In the same
spirit \citet{kersolof99} presented a major survey of CO radio line
emission from irregularly variable (IRV) and semiregularly variable
(SRV) M-type AGB-stars.  They increased the number of IRVs (22
detections), in particular, and SRVs (43 detections) detected in
circumstellar CO emission substantially ($\approx$60\% of the
SRVs and all but one of the IRVs were detected for the first time). 
\citet{youn95} and \citet{groeetal99} have made extensive surveys of
short-period M-Miras.

In this paper we use the radiative transfer method of \citet{schoolof01}
to estimate reasonably accurate mass loss rates and gas expansion
velocities for the \citet{kersolof99} sample.  Comparisons between
these properties and other stellar properties are done, as well as
comparisons with the results for the C-star sample.

\section{The sample}
\label{s:sample}

The original sample consisted of all IRV and SRV AGB-stars of spectral
type M (determined by spectral classification, or using the IRAS LRS
spectra) in the General Catalogue of Variable Stars [GCVS;
\citet{kholetal90}] with an IRAS quality flag 3 in the 12, 25, and
60\,$\mu$m bands.  From this sample we selected for the CO radio line
observations objects with IRAS 60$\mu$m fluxes, $S_{60}$, typically
$\gtrsim$3\,Jy.  Since $S_{60}$ for a star with a luminosity of
4000\,L$_{\odot}$ (see below) and a temperature of 2500\,K is 34\,Jy,
6\,Jy, and 1.4\,Jy at a distance of 100\,pc, 250\,pc, and 500\,pc,
respectively, we added a colour selection criterium.  Only stars
redder than 1.2 mag in the IRAS [12]--[25] colour were observed, thus
biasing the sample towards stars with detectable circumstellar dust
envelopes.  There is a possibility that stars with detectable gas mass
loss rates, but with very little circumstellar dust, were missed due
to this.  About 50\% of the objects, i.e., 109 sources, were subsequently
searched for circumstellar radio line emission.

The CO ($J$=1$\rightarrow$0, 2$\rightarrow$1,
3$\rightarrow$2, and 4$\rightarrow$3) data which are used as the
observational constraints for the mass loss rate determinations in this
paper were presented in \citet{kersolof99}.  They were obtained using
the 20\,m telescope at Onsala Space Observatory (OSO), Sweden, the
15\,m Swedish-ESO Submillimetre Telescope (SEST) on La Silla, Chile,
the IRAM 30\,m telescope on Pico Veleta, Spain, and the James Clerk
Maxwell Telescope on Mauna Kea, Hawaii.  A few additional sources were
observed at OSO in May 2000, see Sect.~\ref{s:observations}.  In
total, 69 stars were detected, 22 IRVs and 47 SRVs, and 45 were
detected in at least two transitions, and 20 in at least three
transitions.  The detection rate was rather high, about 60\%, and the
conclusions drawn in this paper should be representative for the stars
in our sample.

The distances, presented in Table~\ref{t:modelresults}, were estimated using
an assumed bolometric luminosity of 4000\,L$_{\sun}$.  This value was
chosen in agreement with the typical values derived by
\citet{kersetal97} and \citet{bartetal99} for objects with similar
properties.  Morover, the 13 objects in our sample having Hipparcos
parallax errors better than 20\% have a mean luminosity of
4200\,L$_{\odot}$ (with a standard deviation of
1900\,L$_{\odot}$, i.e., consistent with the assumed luminosity and the parallax
uncertainty).  For a statistical study of a large sample of stars
these distance estimates are adequate (and they were used also for
stars with reliable Hipparcos distances to avoid systematic
differences), although the distance estimate for an individual star
has a rather large uncertainty.

The apparent bolometric fluxes were obtained by integrating the
spectral energy distributions ranging from the visual data over the
near-infrared to the IRAS-range \citep{kershron96,kers99}.

\section{Observations}
\label{s:observations}

\begin{figure*}
       {\includegraphics[width=17.0cm]{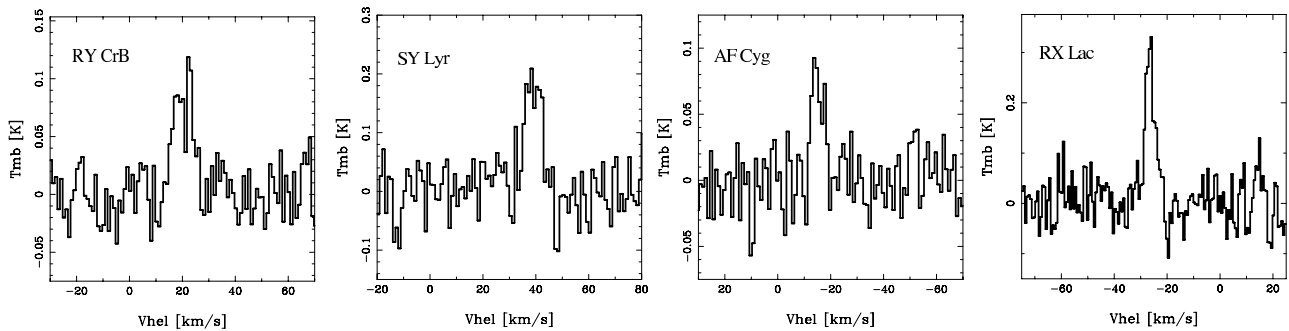}}
        \caption{CO($J$=1$\rightarrow$0) detections at OSO in May 2000}
        \label{f:spectra}
\end{figure*}

A few additional sources were observed at OSO in May 2000 with the
same instrumental setup as used by \citet{kersolof99}.  Relevant
information on the instrumental setup, and the method used to derive
the line profile properties and the upper limits can be found in
\citet{kersolof99}.  A summary of the observational results are given
in Table~\ref{t:obsdata} (where we give the velocity-integrated
intensities, $I_{\rm mb}$, and antenna temperatures, $T_{\rm mb}$,
in main beam brightness scale, the stellar velocities as heliocentric,
$v_{\rm hel}$, and Local Standard of Rest, $v_{\rm LSR}$, velocities,
and the gas expansion velocity, $v_{\rm e}$), and the detections are
presented in Fig.~\ref{f:spectra}.

\begin{table*}[htbp]
  \caption[]{CO($J$=1$\rightarrow$0) results at OSO in May 2000}
  \label{t:obsdata}
  \begin{tabular}{lllccccccl}
  \hline
  \noalign{\smallskip}
  \multicolumn{1}{l}{GCVS4}            &
  \multicolumn{1}{c}{IRAS}             &
  \multicolumn{1}{c}{Var.}             &
  \multicolumn{1}{c}{$I_{\rm mb}$}     &
  \multicolumn{1}{c}{$T_{\rm mb}$}     &
  \multicolumn{1}{c}{$v_{\rm hel}$}    &
  \multicolumn{1}{c}{$v_{\rm LSR}$}    &
  \multicolumn{1}{c}{$v_{\rm e}$}      & Q$^1$ & C \\
     &  &  &  
  \multicolumn{1}{c}{[K\,km\,s$^{-1}$]}      & 
  \multicolumn{1}{c}{[K]}                    & 
  \multicolumn{1}{c}{[km\,s$^{-1}$]}         &
  \multicolumn{1}{c}{[km\,s$^{-1}$]}         &
  \multicolumn{1}{c}{[km\,s$^{-1}$]}         & &\\
  \hline
CX Cas    &02473$+$6313 & SRa &        &        &      &      &                                                            & 5 & IS-lines \\
DP Ori    &05588$+$1054 & SRb & $<$1.3 &        &      &      &                                                            & 5 \\
Z Cnc     &08196$+$1509 & SRb & $<$0.7 &        &      &      &                                                            & 5 \\    
RT Cnc    &08555$+$1102 & SRb & $<$0.5 &        &      &      &                                                            & 5 \\
SX Leo    &11010$-$0256 & SRb & $<$0.9 &        &      &      &                                                            & 5 \\
AF Leo    &11252$+$1525 & SRb & $<$0.7 &        &      &      &                                                            & 5 \\
AY Vir    &13492$-$0325 & SRb & $<$0.8 &        &      &      &                                                            & 5 \\
RY CrB    &16211$+$3057 & SRb & {\phantom{$<$}}0.81 & 0.084              & {\phantom{$-$}}20.4 & {\phantom{$-$}}39.0 & 6.1 & 3 \\
CX Her    &17086$+$2739 & SRb & $<$0.5 &        &      &      &                                                            & 5 \\
IQ Her    &18157$+$1757 & SRb & $<$0.5 &        &      &      &                                                            & 5 \\
V988 Oph  &18243$-$0352 & SRb & $<$0.6 &        &      &      &                                                            & 5 \\
V585 Oph  &18247$+$0729 & SRb & $<$1.4 &        &      &      &                                                            & 5 \\
SY Lyr    &18394$+$2845 & SRb & {\phantom{$<$}}1.4  & 0.18{\phantom{0}}  & {\phantom{$-$}}39.1 & {\phantom{$-$}}58.9 & 4.5 & 2 \\
MZ Her    &18460$+$1903 & SRb & $<$0.6 &        &      &      &                                                            & 5 \\
V858 Aql  &19267$+$0345 & Lb  & $<$2.8 &        &      &      &                                                            & 5 \\
AF Cyg    &19287$+$4602 & SRb & {\phantom{$<$}}0.42 & 0.082              & $-$15.2             & {\phantom{$-$1}}1.6 & 4.2 & 3 \\
V1172 Cyg &19562$+$3304 & Lb  &        &        &      &      &                                                            & 5 & IS-lines   \\
V590 Cyg  &21155$+$4529 & Lb  &        &        &      &      &                                                            & 5 & IS-lines   \\
V655 Cyg  &21420$+$4746 & SRa &        &        &      &      &                                                            & 5 & IS-lines   \\
RX Lac    &22476$+$4047 & SRb & {\phantom{$<$}}1.3  & 0.28{\phantom{0}}  & $-$26.5             & $-$15.8             & 3.5 & 2 \\
   \hline
   \noalign{\smallskip}
   \noalign{$^1$ Quality parameter: 5 (non-det.), 4 (tent. det.), 3 (det., low S/N), 2 (det., good S/N), 1 (det., high S/N)} \\
  \end{tabular}
\end{table*}

\section{Radiative transfer}

We have chosen to use the Monte Carlo method to determine the
excitation of the CO molecules as a function of distance from the central star
\citep{bern79}.  This is a versatile method, close to the physics,
which has been applied to CO radio and far-IR line emission
\citep{crosment97,rydeetal99b,schoetal02}, CO near-IR line scattering
\citep{rydescho01}, and to the OH 1612\,MHz maser emission
\citep{vanlspaa93}.  Sch{\"o}ier \& Olofsson (2000, 2001) applied it to CO radio line data
of samples of AGB carbon stars, and the results suggest that the
modelling of circumstellar CO emission is one of the most reliable
methods for estimating mass loss rates.  The code has been benchmark
tested together with a number of other radiative transfer codes
\citep{vanzetal02}. Below we give a brief introduction to the modelling work, while for
details we refer to \citet{schoolof01}.

\subsection{The CO molecule}

In the excitation analysis of the CO molecule we used 40 rotational
levels in each of the ground and first excited states.  Excitation to
the $v$$=$2 state can be ignored, since the $v$$=$1 state is not well
populated.  The radiative transition probabilities and energy levels were
taken from \citet{chandetal96}.  The collisional rate coefficients
(CO$-$para-H$_2$) for rotational transitions are based on the results
in \citet{flowlaun85}.  These are further extrapolated for $J$$>$11 and
for temperatures higher than 250\,K. We neglect collisional
transitions between the vibrational states because of the low
densities and the relatively low temperatures.

Recently, \citet{flow01} presented revised and extended collisional
rates for CO$-$H$_2$.  Individual rates are generally different from
those previously published in \citet{flowlaun85}, with discrepancies
as large as a factor of two in some cases.  In addition, for
temperatures above 400\,K the rates from \citet{schietal85} were used
and further extrapolated to include transitions up to $J$$=$40.  To
test the effects of the adopted set of collisional rates a number of
test cases with the new rates assuming an H$_2$ ortho-to-para ratio of
three were run.  We found that for the relatively low mass loss rate stars of
interest here, where excitation of CO from radiation dominates over
that from collisions with H$_2$, the adopted set of collisional rates
is only of minor importance.

\subsection{The circumstellar model}

\subsubsection{The geometry and kinematics}

We have adopted a relatively simple, yet realistic, model for the geometry
and kinematics of the CSE. It is assumed that the mass loss is
isotropic and constant with time.  As will be shown below the
median mass loss rate is 2$\times$10$^{-7}$\,M$_{\odot}$\,yr$^{-1}$ and the median gas
expansion velocity is 7.0\,km\,s$^{-1}$ for our sample.  This results
in an extent of the CO envelope of about 2$\times$10$^{16}$\,cm (see
Sect.~\ref{s:coabu}), and the corresponding time scale is about
1000\,yr.  There is now some evidence for mass loss rate modulations
of AGB-stars on this time scale \citep{maurhugg00,mareetal01a}, and
therefore the assumption of a constant mass loss rate may be
questionable, and this must be kept in mind when interpreting the
results. The gas expansion velocity, $v_{\rm e}$, is assumed to be
constant with radius.  This ignores the more complex situation in the
inner part of the CSE, but the emission in the CO lines detected by
radio telescopes mainly comes from the external part.  We further
assume that the hydrogen is in molecular form in the region probed by
the CO emission \citep{glashugg83}.  As a consequence of these
assumptions the H$_2$ number density follows an $r^{-2}$-law.

In the case of low mass loss rate objects the inner radius of the
CSE will have an effect on the model intensities.  The reason is that
radiative excitation plays a role in this case and the absorption of
pump photons at 4.6\,$\mu$m depends (sensitively) on this choice.  We
set the inner radius to 1$\times$10$^{14}$\,cm ($\sim$3\,R$_*$), i.e.,
generally beyond both the sonic point and the dust condensation
radius.  The uncertainty in the mass loss rate estimate introduced by this
assumption is discussed in Sect.~\ref{s:dependence}.  Strictly,
speaking the assumption of a constant expansion velocity from this
inner radius is very likely not correct.  An acceleration region will
enhance the radiative excitation and hence may have an effect on the
estimated mass loss rate.  However, as will be shown in
Sect.~\ref{s:dependence}, the dependence on the inner radius is rather
modest, suggesting that the properties of the inner CSE are not
crucial for the mass loss rate determination.  We have therefore
refrained from introducing yet another parameter, i.e., a velocity law
parameter.

In addition to thermal broadening of the lines microturbulent motions
contribute to the Doppler broadening of the local line width.  We
assume a turbulent velocity width, $v_{\rm t}$, of 0.5\,km\,s$^{-1}$
throughout the entire CSE, i.e., the same value as used by
\citet{schoolof01} [for reference the thermal width of CO is about
0.3$(T_{\rm k}/100)^{0.5}$\,km\,s$^{-1}$].  This can be an important parameter
for low mass loss rate objects since it affects the radial optical
depths and hence the effectiveness of the radiative excitation.  The
constraints on this parameter are rather poor.  The most thorough
analysis in this connection is the one by \citet{huggheal86}.  They
modelled in detail the circumstellar CO line self-absorption in the
high mass loss rate carbon star IRC+10216 and derived a value of
0.9\,km\,s$^{-1}$.  In Sect.~\ref{s:dependence} we discuss the
uncertainty in the mass loss rate estimate introduced by this
parameter.

\subsubsection{Heating and cooling processes}

We determine the kinetic gas temperature in the CSE by taking into
account a number of heating and cooling processes \citep{groe94a}.  The
primary heating process is the viscous heating due to the dust
streaming through the gas medium.  A drift velocity between the gas
and the dust is calculated assuming a dust-driven wind \citep{kwok75}, 
but for the low mass loss rate stars in this study the radiation pressure
on dust may not be very efficient, i.e., the driving of the gas may be
due to something else.  However, as will be explained below, the gas-dust
heating term is nevertheless very uncertain, and we use it as a free
parameter in our model.  Additional heating is due to the
photoelectric effect, i.e., heating by electrons ejected from the grains by
cosmic rays \citep{huggetal88}, but for our low mass loss rate stars
this has a negligible effect inside the CO envelope.

There are three major cooling processes, adiabatic expansion of the
gas, CO line cooling, and H$_2$O line cooling.  The CO line cooling is
taken care of self-consistently by calculating its magnitude after
each iteration using the expression of \citet{crosment97}.  H$_2$O line
cooling is estimated using the results from \citet{neufkauf93}.  They
calculated the H$_2$O excitation using an escape probability method
and estimated the radiative cooling rates for a wide range of
densities and temperatures.  The H$_2$O abundance is set to
2$\times$10$^{-4}$ and the envelope sizes used are based on the
results of \citet{netzknap87}.  In addition, H$_2$ line cooling is
taken into account \citep{groe94a}, but this has negligible effect in
the regions of interest here.

When solving the energy balance equation a number of (uncertain)
parameters describing the dust are introduced.  Following
\citet{schoolof01} we assume that the $Q_{\rm p,F}$-parameter, i.e., the
flux-averaged momentum transfer effeciency from the dust to the gas,
is equal to 0.03 [see \citet{habietal94} for details], and define a
new parameter which contains the other dust parameters,
\begin{equation}
h = \left[\frac{\psi}{0.01}\right]
\left[\frac{2.0\,\mathrm{g\,cm}^{-3}}{\rho_\mathrm{gr}}\right]
\left[\frac{0.05\,\mu\mbox{m}}{a_\mathrm{gr}}\right],
\label{h}
\end{equation}
where $\psi$ is the dust-to-gas mass ratio, $\rho_\mathrm{gr}$ the dust grain
density, and $a_\mathrm{gr}$ its radius.  The normalized values are the
ones used to fit the CO radio line emission of \object{IRC+10216} using this
model \citep{schoolof01}, i.e., $h$=1 for this object.

\subsubsection{The CO fractional abundance distribution}
\label{s:coabu}

We assume that the initial fractional abundance of CO with respect
to H$_2$, $f_{\rm CO}$, is 2$\times$10$^{-4}$, which is the same value
as used by \citet{kahajura94} in their analysis of CO radio line
emission from M-stars.  This is essentially a free parameter, although
its upper limit is given by the abundance of C (i.e.,
7$\times$10$^{-4}$ is the upper limit for a solar C abundance).  Due
to photodissociation by the interstellar radiation field the CO
abundance starts to decline rapidly at a radius, which, for not too
low mass loss rates, depends on the mass loss rate.  Calculations,
taking into account dust-, self- and H$_2$-shielding, and chemical
fractionation, have been performed by \citet{mamoetal88} and
\citet{dotyleun98}.  Here we use the results of \citet{mamoetal88} in
the way adopted by \citet{schoolof01}.  

An approximate expression for
the photodissociation radius, $r_{\rm p}$, consists of two terms, the
unshielded size due to the expansion, which is independent of the mass
loss rate, and the size due to the self-shielding, which scales
roughly as $(f_{\rm CO} \dot{M})^{0.5}$ \citep{schoolof01}.  These
terms are equal at a mass loss rate of about
4$\times$10$^{-8}$$(v_{\rm e}/7)^2$\,M$_{\odot}$\,yr$^{-1}$ for the
adopted CO abundance ($v_{\rm e}$ is given in km\,s$^{-1}$).  That is,
self-shielding plays a role for essentially all of our objects.  We
note that for low mass loss rate objects the spatial extent of the CO
envelope is particularly important since the spatial extent of the CO
line emission is limited by this, and not by excitation, Sect.~\ref{s:dependence}.

\begin{table*}
\caption[ ]{The effect on the velocity-integrated model intensities (in
percent), due to changes in various parameters.  Three model stars
with nominal mass loss rate and gas expansion velocity characteristics
typical for our sample are used.  They lie at a distance of 250\,pc,
and have luminosities of 4000\,L$_{\odot}$ and blackbody temperatures
of 2500\,K. The nominal CSE parameters are $h$=0.2, $r_{\rm
i}$=2$\times$10$^{14}$\,cm, $v_{\rm t}$=0.5\,km\,s$^{-1}$, and $f_{\rm
CO}$=2$\times$10$^{-4}$.  The CO($J$=1$\rightarrow$0),
CO($J$=2$\rightarrow$1), and CO($J$=3$\rightarrow$2) lines are
observed with beam widths of 33$\arcsec$, 23$\arcsec$, and
14$\arcsec$, respectively.  The model integrated line intensities,
$I$, are given for the nominal parameters}
\begin{flushleft}
\begin{tabular}{crrrrrrrrrrrrr}
\hline
\noalign{\smallskip}
  & & &
 \multicolumn{3}{c}{4$\times$10$^{-8}$\,M$_{\odot}$\,yr$^{-1}$, 5\,km\,s$^{-1}$} & &
 \multicolumn{3}{c}{1$\times$10$^{-7}$\,M$_{\odot}$\,yr$^{-1}$, 7\,km\,s$^{-1}$} & &
 \multicolumn{3}{c}{5$\times$10$^{-7}$\,M$_{\odot}$\,yr$^{-1}$, 10\,km\,s$^{-1}$} 
 \\ 
 \cline{4-6}
 \cline{8-10}
 \cline{12-14}
 \multicolumn{1}{c}{Parameter} &
 \multicolumn{1}{c}{Change} & &
 \multicolumn{1}{r}{1$-$0} & 
 \multicolumn{1}{r}{2$-$1} &
 \multicolumn{1}{r}{3$-$2} & 
 & 
 \multicolumn{1}{r}{1$-$0} &
 \multicolumn{1}{r}{2$-$1} & 
 \multicolumn{1}{r}{3$-$2} & 
 &
 \multicolumn{1}{r}{1$-$0} & 
 \multicolumn{1}{r}{2$-$1} &
 \multicolumn{1}{r}{3$-$2} 
 \\
\noalign{\smallskip}
\hline
\noalign{\smallskip}
$I$\,[K\,km\,s$^{-1}$] &    && 0.043  & 0.39   & 1.80   && 0.25   & 1.53   & 4.40  && 2.64   & 6.61   & 13.00 \\
\hline
\noalign{\smallskip}
$\dot{M}$       & $-$50\%   && $-$75  & $-$75  & $-$70  && $-$80  & $-$75  & $-$60 && $-$55  & $-$50  & $-$45 \\
                & $+$100\%  && $+$530 & $+$300 & $+$150 && $+$360 & $+$140 & $+$75 && $+$140 & $+$90  & $+$75 \\
$L$             & $-$50\%   && $+$80  & $+$40  & 0      && $+$70  & $-$5   & $-$30 && $-$15  & $-$35  & $-$40 \\
                & $+$100\%  && $-$35  & $-$30  & $-$20  && $-$50  & $-$20  & $+$10 && $+$10  & $+$30  & $+$35 \\
$h$             & $-$50\%   && $+$5   &	$+$10  & $+$5   && $+$30  & $+$5   & $-$5  && $-$5   & $-$20  & $-$30 \\
                & $+$100\%  && $-$5   &	$-$10  & $-$5   && $-$15  & $-$10  & 0     && $-$5   & $+$15  & $+$30 \\
$r_{\mathrm p}$ & $-$50\%   && $-$75  & $-$75  & $-$65  && $-$80  & $-$70  & $-$50 && $-$60  & $-$35  & $-$20 \\
                & $+$100\%  && $+$300 & $+$170 & $+$80  && $+$250 & $+$80  & $+$35 && $+$60  & $+$15  & $+$5 \\
$r_{\mathrm i}$ & $-$50\%   && $+$15  & $+$20  & $+$10  && $+$25  & 0      & $-$10 && $-$5   & $-$5   & 0 \\
	        & $+$100\%  && $-$10  & $-$10  & $-$10  && $-$15  & $-$5   & 0     && $+$5   & $+$5   & 0 \\
$v_{\mathrm t}$ & $-$50\%   && $+$15  & $+$25  & $+$5   && $+$40  & 0      & $-$15 && $-$5   & $-$5   & $-$5 \\
	        & $+$100\%  && $-$20  & $-$20  & $-$10  && $-$30  & $-$15  & $+$5  && $+$15  & $+$15  & $+$15 \\
\noalign{\smallskip}
\hline
\noalign{\smallskip}
\end{tabular}
\end{flushleft}
\label{t:pardep}
\end{table*}

\subsubsection{The radiation fields}
\label{s:radiation}

The radiation field is provided by two sources.  The central radiation
emanates from the star, and was estimated from a fit to the spectral
energy distribution (SED), usually by assuming two blackbodies, one
representing the direct stellar radiation and one the dust-processed
radiation \citep{kershron96}.  The dust mass loss rates of our sample
stars are low enough that the latter can be ignored.  The temperatures
derived are given in Table~\ref{t:modelresults}.  The stellar
blackbody temperature $T_{\rm bb}$ derived in this manner is generally
about 500\,K lower than the effective temperature of the star
\citep{kershron96}.  The second radiation field is provided by the
cosmic microwave background radiation at 2.7\,K.

\subsection{Dependence on parameters}
\label{s:dependence}

We have checked the sensitivity of the calculated
intensities on the assumed parameters for a set of model stars. 
The model stars have nominal mass loss rate and gas expansion
velocity combinations which are characteristic of our sample:
(4$\times$10$^{-8}$\,M$_{\odot}$\,yr$^{-1}$, 5\,km\,s$^{-1}$),
(1$\times$10$^{-7}$\,M$_{\odot}$\,yr$^{-1}$, 7\,km\,s$^{-1}$), and
(5$\times$10$^{-7}$\,M$_{\odot}$\,yr$^{-1}$, 10\,km\,s$^{-1}$).  They
are placed at a distance of 250\,pc (a typical distance of our stars),
and the nominal values of the other parameters are
$L$=4000\,L$_{\odot}$, $T_{\rm bb}$=2500\,K, $h$=0.2 (the value
adopted for the majority of our stars, Sect.~\ref{s:h}), $r_{\rm
i}$=2$\times$10$^{14}$\,cm (this is twice the inner radius used in the
modelling), $v_{\rm t}$=0.5\,km\,s$^{-1}$, $f_{\rm
CO}$=2$\times$10$^{-4}$, and $r_{\rm p}$ is calculated from the
photodissociation model (Sect.~\ref{s:coabu}).  The
CO($J$=1$\rightarrow$0), CO($J$=2$\rightarrow$1), and
CO($J$=3$\rightarrow$2) lines are observed with beam widths of
33$\arcsec$, 23$\arcsec$, and 14$\arcsec$, respectively.  These are
characteristic angular resolutions of our observations.  Note that, to
some extent, the presented results are dependent on the assumed
angular resolution since resolution effects may play a role.  We
change all parameters (except the expansion velocity) by $-50$\% and
$+100$\% and calculate the velocity-integrated intensities.

The results are summarized in Table~\ref{t:pardep}  in terms of
percentage changes.  Although the dependences are somewhat complicated there
are some general trends.  The line intensities are sensitive to
changes in the mass loss rate, more the lower the mass loss rate, and
hence are sensitive measures of this property.  There is a dependence
on luminosity, in particular for low-$J$ lines for low mass loss rates
and for high-$J$ lines for higher mass loss rates.  The dependence on
the (uncertain) $h$-parameter is fortunately rather weak.  The
dependence on the photodissociation radius is substantial, in
particular for the low-$J$ lines and for low mass loss rates.  The
dependence on the inner radius is weak, and so is the dependence on
the turbulent velocity width.  Thus, we conclude that for our objects
the CO radio line intensities are good measures of the mass loss rate,
but it shall be kept in mind that they are rather dependent on the
uncertain photodissociation radius and, to some extent, on the assumed
luminosity.  A similar sensitivity analysis for C-stars were done by
\citet{schoolof01}, and studies of the parameter dependence were done
by \cite{kast92} and \cite{kwanwebs93}.

There is also a dependence on the adopted CO abundance.  For the low mass
loss rates considered here
($\lesssim$5$\times$10$^{-7}\,M_{\odot}$\,yr$^{-1}$) a constant
product $f_{\rm CO} \dot{M}$ produces the same model line intensities. 
Hence, mass loss rates for a different value of $f_{\rm CO}$ are
easily obtained.  The reason for this behaviour is that the size of
the emitting region is photodissociation limited rather than
excitation limited, i.e., for our objects it scales roughly as
$(f_{\rm CO} \dot{M})^{0.5}$, see Sect.~\ref{s:coabu}.  Furthermore,
the energy levels are to a large extent radiatively excited, i.e., the
density plays less of a role for the excitation.  Therefore, since the
emission is optically thin, a change in the abundance must be
compensated by an equal, but opposite, change in the mass loss rate to
keep the calculated intensities unchanged.

An additional uncertainty is due to the somewhat crude treatment of
H$_2$O line cooling.  The modelling shows that this cooling process
has an effect on the temperature structure of the CSE. However, it
is found to be of importance only in the inner warm and dense part of the
CSE where H$_2$O is abundant.  The size of the H$_2$O envelope is only
about a tenth of the CO envelope.  This small region contributes only
marginally to the line intensities of the observed low-$J$ transitions
(see for instance the weak dependence on $r_{\rm i}$).

\begin{figure*}
       {\includegraphics[width=17.0cm]{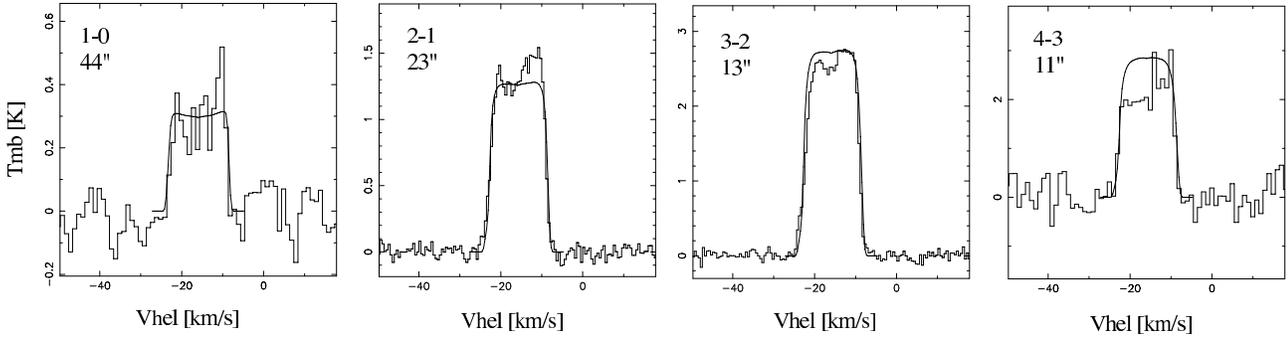}} 
       \caption{\object{SW~Vir}
       CO($J$=1$\rightarrow$0 and 2$\rightarrow$1) spectra obtained with the
       SEST, and CO($J$=3$\rightarrow$2 and 4$\rightarrow$3) spectra obtained
       with the JCMT (histograms).  The line profiles from the best-fit model
       are shown as solid, thin lines (the beam size is given in each panel)}
        \label{f:swvir}
\end{figure*}

\section{Model results}

\subsection{Best-fit model strategy}

The radiative transfer analysis produces model brightness
distributions.  These are convolved with the appropriate beams to
allow a direct comparison with the observed velocity-integrated line
intensities and to search for the best fit model.  There are two
remaining parameters to vary in this fitting procedure, the mass loss
rate $\dot{M}$ and the $h$-parameter.  These two parameters were
varied until the best-fit model was found.  The
quality of a fit was quantified using the chi-square statistic,
\begin{equation}
\label{chi2_eq}
\chi^2_{\rm red} = \frac{1}{\rm N-p}\,\sum^{\rm N}_{i=1} \frac{[I_{\mathrm{mod},i} -
I_{\mathrm{obs},i})]^2}{\sigma^2_{i}},
\end{equation}
where $I$ is the total integrated line intensity, $\sigma_i$ the
uncertainty in observation $i$, p the number of free parameters, and the
summation is done over all independent observations N. The errors in
the observed intensities are always larger than the calibration
uncertainty of $\sim$20\%.  We have chosen to adopt
$\sigma_i$\,=\,0.2$I_{\mathrm{obs},i}$ to put equal weight on all
lines, irrespective of the S/N-ratio. Initially a grid, centered
on $\dot{M}$=10$^{-7}$\,M$_{\odot}$\,yr$^{-1}$ and $h$=0.1, with step sizes of
50\% in $\dot{M}$ and 100\% in $h$ was used to locate the
$\chi^2$-minimum.  The final parameters were obtained by decreasing
the step size to 25\% in $\dot{M}$ and 50\% in $h$ and by interpolating
between the grid points. The final chi-square values for stars
observed in more than one transition are given in
Table~\ref{t:modelresults}.  The line profiles were not used to
discriminate between models, but differences between model and
observed line profiles are discussed in Sect.~\ref{s:lineprof}.

In general, the model results fit rather well the observed data as can be
seen from the chi-square values.  For instance, we reproduce the very
high ($J$=2$\rightarrow$1)/($J$=1$\rightarrow$0) intensity ratios
reported for these objects, 4.2 on average and ratios of 10 are not
uncommon \citep{kersolof99}.  Table~\ref{t:pardep} gives the
integrated line intensities of our model stars in
Sect~\ref{s:dependence}.  These give an indication of how the line
intensities depend on the mass loss rate.  In particular, one should
note the large intensity ratios for low mass loss rates:
\mbox{$I(2-1)/I(1-0)$} equals about 9, 5, and 3, and
\mbox{$I(3-2)/I(1-0)$} equals about 42, 13, and 5 for
4$\times$10$^{-8}$\,M$_{\odot}$\,yr$^{-1}$,
1$\times$10$^{-7}$\,M$_{\odot}$\,yr$^{-1}$, and
5$\times$10$^{-7}$\,M$_{\odot}$\,yr$^{-1}$, respectively.  In
Fig.~\ref{f:swvir} we present the observational data of \object{SW~Vir} and the
best-fit model results.

\subsection{The h-parameter}
\label{s:h}

\begin{figure}
\centering
	{\includegraphics[width=7.0cm]{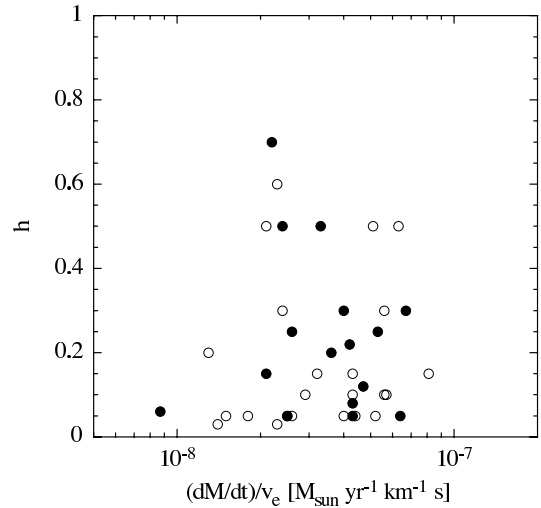}}
        \caption{The $h$-parameters derived from the radiative transfer
        analysis plotted against the density measure $\dot{M}/v_{\rm e}$. 
        Objects with three or more transitions observed are marked with filled
        circles, while those with only two transitions observed are marked
        with open circles}
        \label{f:hmdotve}
\end{figure}

The intensity ratios between lines of different excitation
requirements are sensitive to the temperature structure.  Therefore,
we initially used stars with three, or more, transitions observed to
estimate $h$.  In total, 16 objects fulfil this criterium and the
resulting model fits are rather good as shown by the $\chi^2_{\rm
red}$-values, Table~\ref{t:modelresults}.  The derived $h$-values have a
mean of 0.24 and a median of 0.21.  The scatter in the derived
$h$-values is rather large, and there is no apparent trend with the
density measure $\dot{M}/v_{\rm e}$, Fig.~\ref{f:hmdotve}.  We also
determined, in the same way, the $h$-values for those stars with only
two observed transitions (25 objects), and the result was a mean of
0.22 and a median of 0.1.  There is no trend with the density measure
for these objects either, and the scatter is large,
Fig.~\ref{f:hmdotve}.  As outlined above the line intensities of low
mass loss rate objects are not particularly sensitive to $h$, and this
very likely contributes to the large scatter in the derived values. 
The median value is clearly lower for the sources observed in only two
transitions.  This is probably due to a systematic effect.  Pointing
and calibration problems tend to affect more the higher-frequency
lines, and will, on average, lead to too low observed intensities.  A
low intensity ratio between a higher-frequency and a lower-frequency
line can be accomodated in the model only by ``cooling'' the envelope,
i.e., by lowering $h$.  A decrease in $h$ must be compensated by an
increase in $\dot{M}$ to preserve the line intensities.  To avoid this
systematic effect we used $h$=0.2 for all objects observed in less
than three transitions.  We conclude that the derived mass loss rates
are not sensitively dependent on this choice, unless the true $h$ is
considerably lower than this value.  We note from the results of
\citet{schoolof01} that an increase in $h$ will lead to a decrease in
$\dot{M}$, and vice versa.

\citet{schoolof01} found an $h$-value of about 1 for the high mass loss
rate carbon stars, and a trend of decreasing values towards lower mass
loss rates, reaching an average of about 0.2 in the mass loss rate
range of our stars. The presumed difference in grain density between
carbon grains (2.0\,g\,cm$^{-3}$) and silicate grains
(3.0\,g\,cm$^{-3}$) means that for the same grain size our $h$-value
of 0.2 indicates a dust-to-gas mass ratio which is 1.5 times higher,
i.e., 3$\times$10$^{-3}$, in the CSEs of M-type stars.  However, the
uncertainties are such that we can only conclude that both the M-type
and C-type CSEs due to low mass loss rates appear to have dust
properties significantly different from those of C-type CSEs due to
high mass loss rates (however, all mass loss rate determinations are
made using the same value for the flux-averaged dust momentum transfer
efficiency, which determines the gas-dust drift velcoity and hence
affects the heating of the CSE, while in reality it may depend on the
mass loss rate).  We can also conclude that the gas-CSEs due to low mass loss
rates are cooler than expected from a simple extrapolation of the
results for \object{IRC+10216}
\citep{groeetal98b,crosment97,schoolof00}.

\begin{table*}
    \centering
    \caption[]{Source parameters and model results}
     \label{t:modelresults}
   \[
	\begin{tabular}{llrrcccclrc} 
\hline
\noalign{\smallskip}
Source & Var.  & P{\phantom{00}} & D$^1$ & T$_{\rm bb}$ & $\dot{M}$ & $v_{\rm e}$ & $r_{\rm p}$ & $h$ & $\chi^2_{\rm red}$ & N \\
       & type & [days] & [pc] & [K] & [10$^{-7}$\,M$_{\odot}$ yr$^{-1}$] & [km s$^{-1}$] & [10$^{16}$\,cm] & & & \\
\hline
BC And   & Lb  &     & 450 & 2510 &  2.0             & {\phantom{1}}4.0 & 2.0 & & & 1 \\
CE And   & Lb  &     & 740 & 2720 &  5{\phantom{.0}} &             10.5 & 2.5 & & & 1 \\
RS And   & SRa & 136 & 290 & 2620 &  1.5             & {\phantom{1}}4.4 & 1.6 & & 0.7 & 2 \\
UX And   & SRb & 400 & 280 & 2240 &  4{\phantom{.0}} &             12.8 & 2.1& & 1.9  & 2 \\
TZ Aql   & Lb  &     & 470 & 2460 &  1.0             & {\phantom{1}}4.8 & 1.3 & & & 1 \\
V584 Aql & Lb  &     & 390 & 2340 &  0.5             & {\phantom{1}}2.2 & 1.2 & & & 1 \\
AB Aqr   & Lb  &     & 460 & 2580 &  1.3             & {\phantom{1}}4.2 & 1.5 & & & 1 \\
SV Aqr   & Lb  &     & 470 & 2180 &  3{\phantom{.0}} & {\phantom{1}}8.0 & 2.1 & & 9.1 & 2\\
$\theta$ Aps & SRb & 119 & 110 & 2620 & 0.4          & {\phantom{1}}4.5 & 0.8 & & 9.8 & 2\\
T Ari    & SRa & 317 & 310 & 2310 &  0.4             & {\phantom{1}}2.4 & 0.9 & & & 1 \\
RX Boo   & SRb & 340 & 110 & 2220 &  5{\phantom{.0}} & {\phantom{1}}9.3 & 2.6 & & 1.4 & 2 \\
RV Cam   & SRb & 101 & 350 & 2570 &  2.5             & {\phantom{1}}5.8 & 2.0 & & 0.4 & 2 \\
BI Car   & Lb  &     & 430 & 2420 &  0.3             & {\phantom{1}}2.2 & 0.9 & & & 1 \\
V744 Cen & SRb & 90  & 200 & 2750 &  1.3             & {\phantom{1}}5.3 & 1.5 & 0.05 & 6.0 & 3 \\
SS Cep   & SRb & 90  & 340 & 2580 &  6{\phantom{.0}} &             10.0 & 2.7 &     & 3.3 & 2\\
UY Cet   & SRb & 440 & 300 & 2400 &  2.5             & {\phantom{1}}6.0 & 2.0 & 0.2 & 0.4 & 3 \\
CW Cnc   & Lb  &     & 280 & 2400 &  5{\phantom{.0}} & {\phantom{1}}8.5 & 2.5 & 0.25 & 3.2 & 3 \\
RY CrB   & SRb &     & 550 & 2340 &  4{\phantom{.0}} & {\phantom{1}}5.7 & 2.5 & & & 1 \\
R Crt    & SRb & 160 & 170 & 2130 &  8{\phantom{.0}} &             10.6 & 3.0 & 0.3 &  0.7 & 4\\
AF Cyg   & SRb &     & 300 & 2840 &  0.8             & {\phantom{1}}3.5 & 1.2 & & & 1 \\
W Cyg    & SRb & 131 & 130 & 2670 &  1.0             & {\phantom{1}}8.3 & 1.3 & & 0.7 & 2   \\
U Del    & SRb & 110 & 210 & 2720 &  1.5             & {\phantom{1}}7.5 & 1.5 & & 10.5 & 2\\
R Dor    & SRb & 338 & 45  & 2090 &  1.3             & {\phantom{1}}6.0 & 1.4 & 0.7 & 1.6 & 3 \\
AH Dra   & SRb & 158 & 340 & 2680 &  0.8             & {\phantom{1}}6.4 & 1.1 & & & 1 \\
CS Dra   & Lb  &     & 370 & 2580 &  6{\phantom{.0}} &             11.6 & 2.7 & 0.05 & 3.2 & 3  \\
S Dra    & SRb & 136 & 270 & 2230 &  4{\phantom{.0}} & {\phantom{1}}9.6 & 2.2 & 0.3  & 0.5 & 3 \\
SZ Dra   & Lb  &     & 510 & 2580 &  6{\phantom{.0}} & {\phantom{1}}9.6 & 2.7 & & & 1 \\
TY Dra   & Lb  &     & 430 & 2300 &  6{\phantom{.0}} & {\phantom{1}}9.0 & 2.8 & & 1.0 & 2 \\
UU Dra   & SRb & 120 & 320 & 2260 &  5{\phantom{.0}} & {\phantom{1}}8.0 & 2.5 & & 5.0 & 2 \\
g Her    & SRb & 89  & 100 & 2700 &  1.0             & {\phantom{1}}8.4 & 1.3 & & & 1 \\
AK Hya   & SRb & 75  & 210 & 2430 &  1.0             & {\phantom{1}}4.8 & 1.3 & 0.15 & 1.5 & 4 \\
EY Hya   & SRa & 183 & 300 & 2400 &  2.5             &             11.0 & 1.8 & & 1.1 & 2 \\
FK Hya   & Lb  &     & 310 & 2630 &  0.6             & {\phantom{1}}8.7 & 1.0 & & & 1 \\
FZ Hya   & Lb  &     & 330 & 2460 &  2.0             & {\phantom{1}}7.8 & 1.6 & & 0.0 & 2 \\
W Hya    & SRa & 361 & 65  & 2090 &  0.7             & {\phantom{1}}6.5 & 1.0 & & & 1 \\
RX Lac   & SRb &     & 250 & 2450 &  0.8             & {\phantom{1}}2.2 & 1.6 & & & 1  \\
RW Lep   & SRa & 150 & 400 & 2150 &  0.5             & {\phantom{1}}4.4 & 0.9 & & & 1  \\
RX Lep   & SRb & 60  & 150 & 2660 &  0.5             & {\phantom{1}}3.5 & 1.0 & & 0.2 & 2 \\
SY Lyr   & SRb &     & 640 & 2410 &  6{\phantom{.0}} & {\phantom{1}}4.6 & 3.6 & & & 1  \\
TU Lyr   & Lb  &     & 420 & 2470 &  3{\phantom{.0}} & {\phantom{1}}7.4 & 2.1 & & 1.3 & 2 \\
U Men    & SRa & 407 & 320 & 2160 &  2.0             & {\phantom{1}}7.2 & 1.7 & & & 1 \\
T Mic    & SRb & 347 & 130 & 2430 &  0.8             & {\phantom{1}}4.8 & 1.2 & & 1.6 & 2 \\
EX Ori   & Lb  &     & 470 & 2490 &  0.8             & {\phantom{1}}4.2 & 1.3 & & 2.2 & 2 \\
V352 Ori & Lb  &     & 250 & 2560 &  0.5             & {\phantom{1}}8.4 & 0.9 & & & 1 \\
S Pav    & SRa & 381 & 150 & 2190 &  0.8             & {\phantom{1}}9.0 & 1.1 & & & 1 \\
SV Peg   & SRb & 145 & 190 & 2330 &  3{\phantom{.0}} & {\phantom{1}}7.5 & 2.1 & 0.1 & 4.2 & 3 \\
TW Peg   & SRb & 929 & 200 & 2690 &  2.5             & {\phantom{1}}9.5 & 1.8 & 0.25 & 3.5 & 4 \\
V PsA    & SRb & 148 & 220 & 2360 &  3{\phantom{.0}} &             14.4 & 1.9 & & & 1 \\
L$^2$ Pup& SRb & 141 & 85  & 2690 &  0.2             & {\phantom{1}}2.3 & 0.7 & 0.05 & 0.9 & 4 \\
OT Pup   & Lb  &     & 500 & 2630 &  5{\phantom{.0}} & {\phantom{1}}9.0 & 2.6 & & 4.8 & 2 \\
Y Scl    & SRb &     & 330 & 2620 &  1.3             & {\phantom{1}}5.2 & 1.5 & & 2.0 & 2 \\
CZ Ser   & Lb  &     & 440 & 2150 &  8{\phantom{.0}} & {\phantom{1}}9.5 & 3.2 & & 0.2 & 2 \\
$\tau^4$ Ser & SRb & 100 & 170 & 2500 & 1.5          &             14.4 & 1.5 & & & 1  \\
SU Sgr   & SRb & 60  & 240 & 2090 &  4{\phantom{.0}} & {\phantom{1}}9.5 & 2.3 & & 4.7 & 2 \\
UX Sgr   & SRb & 100 & 310 & 2520 &  1.5             & {\phantom{1}}9.5 & 1.5 &  & & 1  \\
V1943 Sgr& Lb  &     & 150 & 2250 &  1.3             & {\phantom{1}}5.4 & 1.4 & 0.5 & 9.2 & 3 \\
V Tel    & SRb & 125 & 290 & 2260 &  2.0             & {\phantom{1}}6.8 & 1.6 &  & 9.5 & 2 \\
\hline
\noalign{\smallskip}
\noalign{$^1$ Distance derived assuming a luminosity of 4000\,L$_{\odot}$}\\
\end{tabular}
        \]
\end{table*}

\addtocounter{table}{-1}
\begin{table*}
\caption[ ]{continued.}
   \[
   \begin{tabular}{llrrcccclrc}
   \hline
   \noalign{\smallskip}
Source & Var. & P{\phantom{00}} & D$^1$ & T$_{\rm bb}$ & $\dot{M}$ & $v_{\rm e}$ & $r_{\rm p}$ & $h$ & $\chi^2_{\rm red}$ & N \\
 & type & [days] & [pc] & [K] & [10$^{-7}$\,M$_{\odot}$ yr$^{-1}$] & [km s$^{-1}$] & [10$^{16}$\,cm] & & &\\
\hline
Y Tel    & Lb  &     & 340 & 2350 &  0.5             & {\phantom{1}}3.5  & 1.0 & & 13.3 & 2 \\
AZ UMa   & Lb  &     & 490 & 2620 &  2.5             & {\phantom{1}}4.5  & 2.1 & & 2.7 & 2 \\
Y UMa    & SRb & 168 & 220 & 2230 &  1.5             & {\phantom{1}}4.8  & 1.7 & 0.5 & 0.9 & 3 \\
SU Vel   & SRb & 150 & 250 & 2380 &  2.0             & {\phantom{1}}5.5  & 1.8 & 0.2 & 2.9 & 3 \\
BK Vir   & SRb & 150 & 190 & 2210 &  1.5             & {\phantom{1}}4.0  & 1.9 & 0.05 & 0.1 & 3\\
RT Vir   & SRb & 155 & 170 & 2110 &  5{\phantom{.0}} & {\phantom{1}}7.8  & 2.7 & 0.05 & 0.5 & 4 \\
RW Vir   & Lb  &     & 280 & 2530 &  1.5             & {\phantom{1}}7.0  & 1.5 & & 1.1 & 2\\
SW Vir   & SRb & 150 & 120 & 2190 &  4{\phantom{.0}} & {\phantom{1}}7.5  & 2.2 & 0.1 & 0.7 & 6 \\
\hline
\noalign{\smallskip}
\noalign{$^1$ Distance derived assuming a luminosity of 4000\,L$_{\odot}$}\\
\end{tabular}
        \]
\end{table*}

\begin{figure*}
\centering
    {\includegraphics[width=17.0cm]{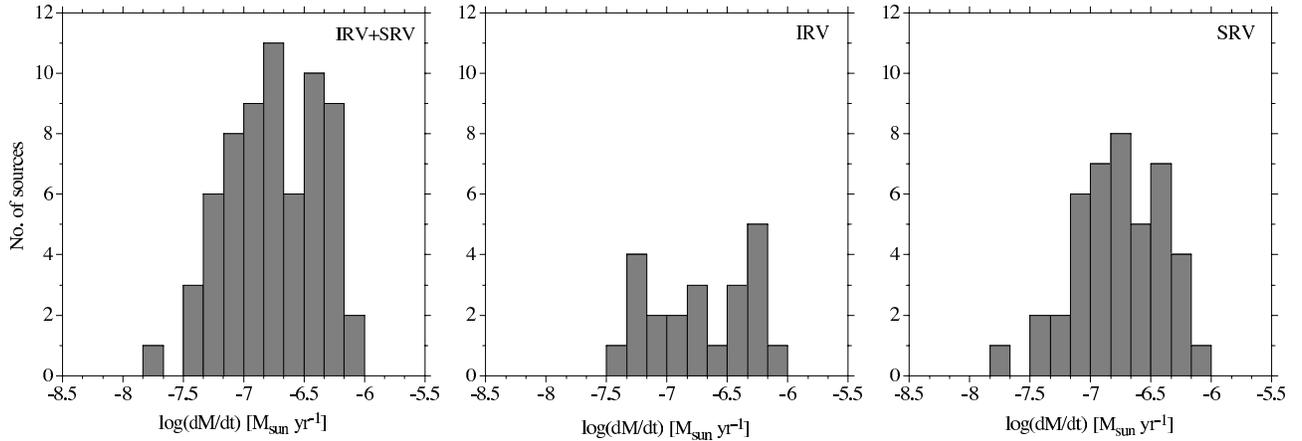}}
        \caption{Mass loss rate distribution
        for the whole sample (left panel), as well as for the IRVs (middle)
        and SRVs (right) separately.  The objects with double-component line
        profiles are excluded}
        \label{f:mldistr}
\end{figure*}

\begin{figure*}
\centering
	{\includegraphics[width=17.0cm]{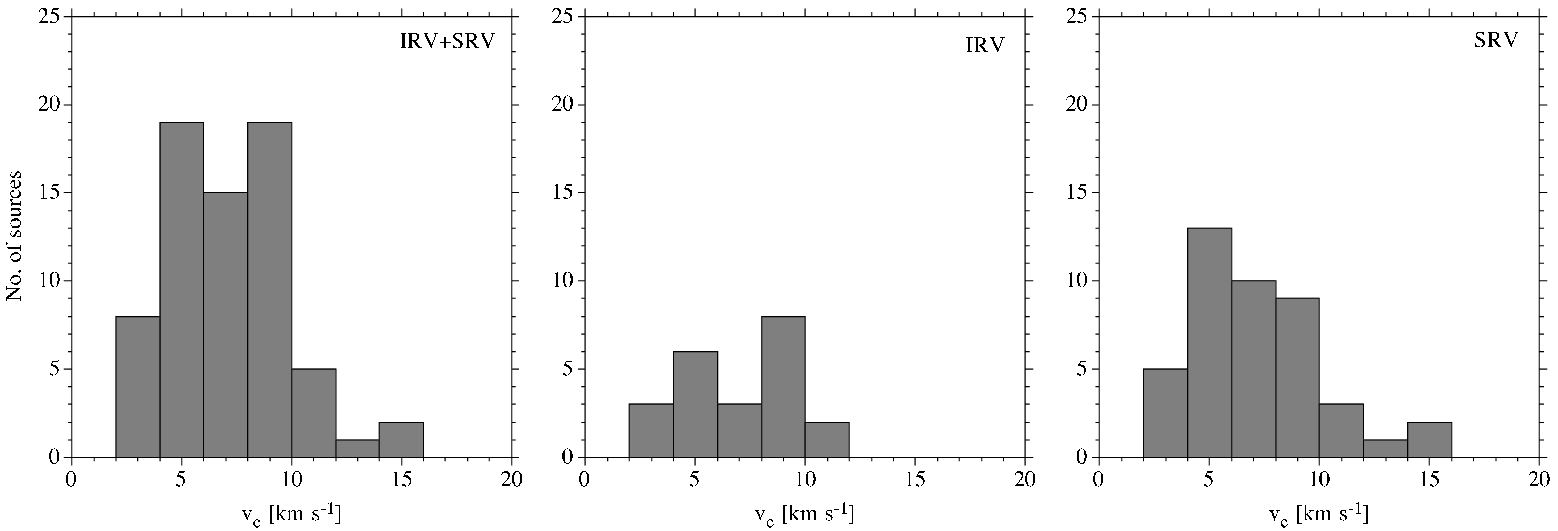}}
        \caption{Gas expansion velocity distribution for the whole sample
        (left panel), as well as for the IRVs (middle) and SRVs (right)
        separately.  The objects with
        double-component line profiles are excluded}
        \label{f:vedistr}
\end{figure*}

\subsection{The mass loss rates}
\label{s:massloss}

The estimated mass loss rates are given in
Table~\ref{t:modelresults} (rounded off to the number nearest to 1.0,
1.3, 1.5, 2.0, 2.5, 3, 4, 5, 6, or 8, i.e., these values are separated
by about 25\%), and their distributions are shown in Fig.~\ref{f:mldistr}
(four sources with clearly peculiar line profiles are discussed
separately in Sect.~\ref{s:doublecomp}).  We estimate
that, within the adopted circumstellar model, the derived mass loss
rates are uncertain by about $\pm$20\% for those sources with three or
more observed transitions, since the line intensities are very
sensitive to changes in this parameter (see Table~\ref{t:pardep}). 
The uncertainty increases when less than three transitions are
observed, generally $\pm$50\%, but it may be as bad as a factor of a
few for objects with low $h$-values.  To this should be added the
uncertainty due to the distance, the luminosity, the photodissociation
radius, the fractional CO abundance, the collisional cross sections,
and the pointing/calibration.  Nevertheless, as a whole, we believe
that these are the most accurate mass loss rates determined for these
types of objects, but on an absolute scale they are uncertain by at
least a factor of a few for an individual object.  A comparison
with the mass loss rates estimated by \citet{knapetal98} for the six
stars in common shows differences by less than a factor of two, except
in the case of \object{RT~Vir} for which we derive a five times higher
value. Note that the mass loss rates given are not corrected for the
He-abundance, i.e., they are molecular hydrogen mass loss rates.

The detailed radiative transfer presented here results in
considerably higher mass loss rates than those obtained with the simpler
analysis in \citet{kersetal96a} and \citet{kersolof98}.  For the 43 objects
in common we derive mass loss rates which are on average ten times
higher for the same distances and CO abundance (the median difference is six). 
This confirms the conclusion by \citet{schoolof01} that the formulae
of \citet{knapmorr85} and \citet{kast92} lead to substantially
underestimated mass loss rates for low mass loss rate objects.  This
is no surprise since both formulae were calibrated against
\object{IRC+10216} (for which $h$=1), and in the case of
\citet{knapmorr85} an older CO photodissociation model was used.

\subsection{The gas expansion velocities}

The gas expansion velocities given in Table~\ref{t:modelresults}
are obtained in the model fits.  Hence, they should be somewhat more accurate
than the pure line profile fit results given in \citet{kersolof99},
since for instance the effect of turbulent broadening is taken into
account.  The former are in general somewhat lower than the latter.  We
estimate the uncertainty to be of the order $\pm$10\%.  The
uncertainty is dominated by the S/N-ratio since the spectral
resolution is in most cases more than adequate.  A significant
fraction of the sources has gas expansion velocities lower than
5\,km\,s$^{-1}$, and for these the assumption of a turbulent velocity
width of 0.5\,km\,s$^{-1}$ will have some effect on the expansion
velocity estimate.  The gas expansion velocity distribution for the whole
sample, as well as those of the IRVs and SRVs separately, are shown in
Fig.~\ref{f:vedistr} (excluding the double-component sources,
Sect.~\ref{s:doublecomp}).

\section{Discussion}

In this section we present and discuss a number of results based on
the derived mass loss rates and gas expansion velocities.  Extensive
comparisons are made with the results of the C-type IRVs and SRVs (39
objects) analysed by \citet{schoolof01} using the same methods as in
this study.

\subsection{The mass loss rate distribution}

The distribution of the derived mass loss rates have a median value of
2.0$\times$10$^{-7}$\,M$_{\odot}$\,yr$^{-1}$, and a minimum of
2.0$\times$10$^{-8}$\,M$_{\odot}$\,yr$^{-1}$ and a maximum of
8$\times$10$^{-7}$\,M$_{\odot}$\,yr$^{-1}$.  There is no significant
difference between the IRVs and the SRVs, but the number of IRVs is
quite low.  We believe that these mass loss rate
distributions are representative for the mass losing M-type IRVs and SRVs on
the AGB (see arguments below).  We find no significant difference when
comparing with the sample of C-type IRVs and SRVs, where the median was
1.6$\times$10$^{-7}$\,M$_{\odot}$\,yr$^{-1}$.  Therefore,
the mass loss rates of these types of variables appear independent of
chemistry (also for the C-stars there is no dependence on the C/O-ratio).  This
conclusion rests on the uncertain assumptions of the CO fractional
abundance (10$^{-3}$ in the case of the C-star sample, as opposed to
the value 2$\times$10$^{-4}$ used here for the M-stars).

Most notable is the sharp cut-off at a mass loss rate slightly below
10$^{-6}$\,M$_{\odot}$\,yr$^{-1}$.  This is very likely not a selection
effect.  All our stars are included in the GCVS, and even though this
favours less obscured stars the GCVS contains many stars with mass
loss rates in excess of this value.  Thus, there appears to be an upper
limit for the mass loss rate of an M-type IRV or SRV on the AGB.
Regular pulsators of M-type, Miras and OH/IR-stars, clearly reach
significantly higher mass loss rates, and hence the regularity of the
pulsation and its amplitude play an important role for the magnitude
of the mass loss rate. Some caution must be exercised here
though.  We are averaging the mass loss rate over a time scale of
about one thousand years, and there are indications that the mass loss
rates of IRV/SRVs are more time-variable, on shorter time scales than
this, than are the mass loss rates of the Miras \citep{mareetal01a}.  This
would lead to an, on the average, lowered mass loss rate of an IRV/SRV.

\begin{figure}
\centering
	{\includegraphics[width=7.0cm]{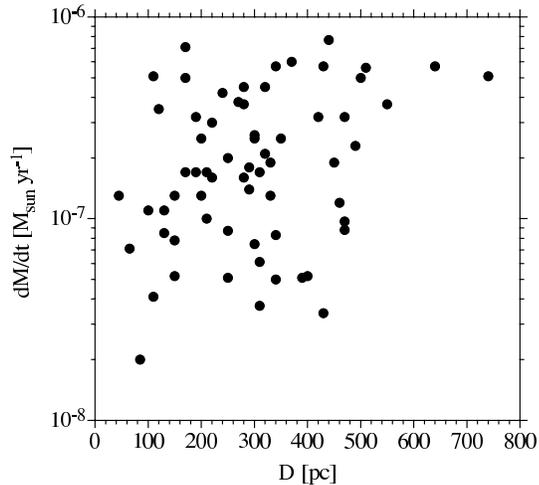}}
        \caption{The derived mass loss rate as a function of the distance
	to the object}
        \label{f:mdotd}
\end{figure}

The decrease in the number of objects with low mass loss rates could
indicate an effect of limited observational sensitivity.  However, a
plot of the mass loss rate as a function of the distance suggests that
this is not the case, Fig.~\ref{f:mdotd}.  We detect low mass loss
rate objects out to about 500\,pc, and beyond this only a few higher
mass loss rate objects are detected.  That is, nearby stars with mass
loss rates below 10$^{-8}$\,M$_{\odot}$\,yr$^{-1}$ should be
detectable.  [We checked all the non-detections reported by
\citet{kersolof99}, and found that in no case do they provide an upper
limit in the mass loss rate which is significantly lower than a few
10$^{-8}$\,M$_{\odot}$\,yr$^{-1}$.]  Hence, we interpret the trailing
off at low mass loss rates as due to a lack of such sources among the
mass losing M-type IRVs and SRVs on the AGB. However, our sample is
limited by the IRAS colour [12]--[25] (Sect.~\ref{s:sample}). 
Therefore, it is possible that there exists M-type AGB-stars with mass
loss rates lower than our limit of about
10$^{-8}$\,M$_{\odot}$\,yr$^{-1}$.  The case for the C-stars is
different.  \citet{schoolof01} also derived a lower limit of about
10$^{-8}$\,M$_{\odot}$\,yr$^{-1}$, but this is based on a K-magnitude
limited sample, for which the K-magnitude is expected to be relatively
constant, where all stars within about 500\,pc were detected
\citep{olofetal93a}.

\begin{figure}
\centering
	{\includegraphics[width=7.0cm]{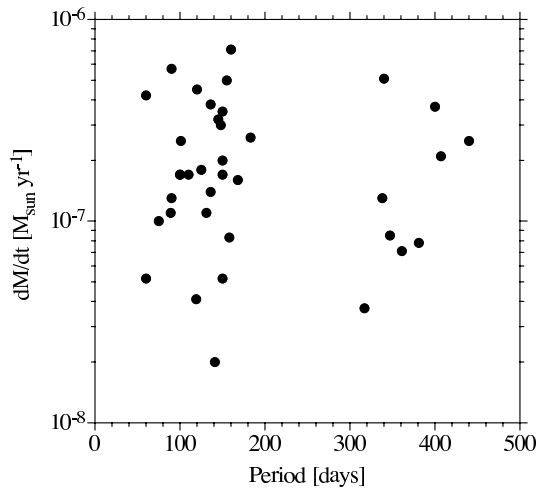}}
	\caption{The derived mass loss rate as a function of the 
	period of pulsation for the SRVs}
        \label{f:mdotp}
\end{figure}

We have also compared the derived mass loss rates with the periods of
the SRVs, Fig.~\ref{f:mdotp}.  The conclusion by \citet{kersetal96a}
that for these objects the period of pulsation plays no role for the
mass loss rate still holds.  The apparent absence of stars with
periods in the range 200--300 days is most probably due to a distinct
division into two pulsational modes, one operating only below 200 days
and one only above 300 days.  Likewise, for the C-SRVs we find no
correlation between mass loss rate and period (the gap between 200
and 300 days does not exist for these stars).

\subsection{Mass loss and stellar temperature}

\begin{figure}
\centering
	{\includegraphics[width=7.0cm]{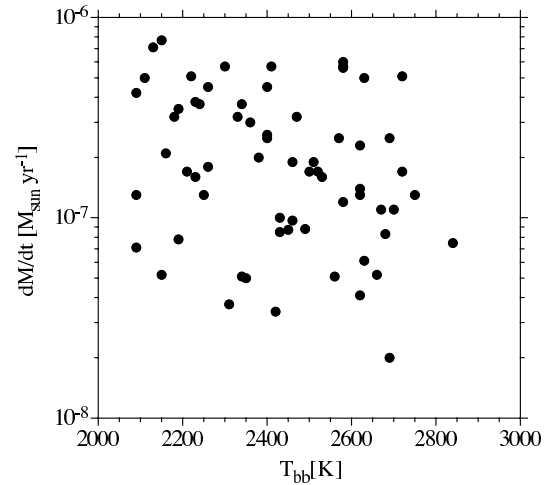}}
	\caption{The derived mass loss rate as a function of the 
	stellar blackbody temperature}
        \label{f:mdottbb}
\end{figure}

It has turned out to be very difficult to derive the mass loss rate of
an AGB-star from first principles.  Some attempts have nevertheless
been made and they all indicate a strong dependence on the stellar
effective temperature, due to its effect on grain condensation
\citep{arndetal97,wintetal00b}.  The blackbody temperatures derived for
our stars are at least indicative of the stellar effective temperatures,
even though there may be a systematic effect (Sect.~\ref{s:radiation}).  Of
course, a high mass loss rate will lead to significant dust emission
and this may have an effect on the stellar blackbody temperature estimate, but
the approach with two blackbodies should to some extent diminish this
effect.  We have in Fig.~\ref{f:mdottbb} plotted the derived mass loss
rate as a function of the stellar blackbody temperature.  Clearly, there is no
correlation at all.  Even taking into account the somewhat uncertain
relation between our stellar blackbody temperature and the effective
temperature this shows that for these objects with relatively low mass
loss rates the temperature in the stellar atmosphere plays no role. 
The C-type IRVs and SRVs show the same absence of a correlation,
only when the C-Miras are added is there a weak trend with mass loss
rate decreasing with increasing temperature.

\subsection{The gas expansion velocity distribution}

The gas expansion velocities derived from the model fits have a
distribution with a median for the whole sample of 7.0\,km\,s$^{-1}$,
and a minimum of 2.2\,km\,s$^{-1}$ and a maximum of
14.4\,km\,s$^{-1}$, i.e., clearly these objects sample the low gas
expansion velocity end of AGB winds.  Similar results have been
obtained for short-period M-Miras \citep{youn95,groeetal99}.  We find no
apparent difference between the IRV and the SRV distributions, but the
former is based on relatively few objects.  A comparison with the
C-type IRVs and SRVs, for which the median is 9.5\,km\,s$^{-1}$ (39
objects) and where the fraction of low-velocity sources is much lower,
indicates that in this respect there is a difference between the
chemistries.  A C-type chemistry produces higher gas expansion
velocities.  The large fraction of low-velocity sources in our sample
is further discussed in Sect~\ref{s:lowve}.

\subsection{Mass loss and envelope kinematics}

\begin{figure}
\centering
	{\includegraphics[width=7.0cm]{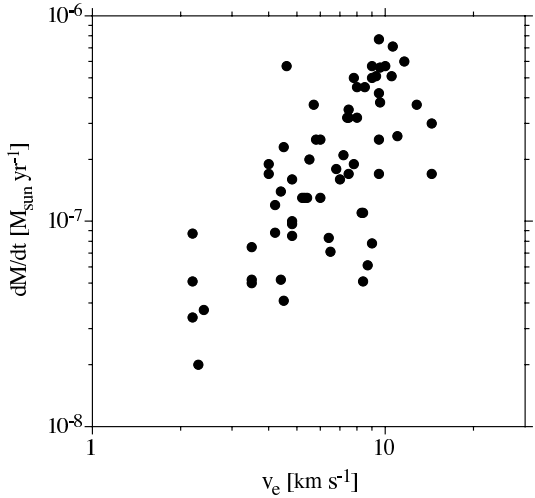}}
        \caption{The derived mass loss rate as a function of the gas
	expansion velocity for the whole sample, excluding the 
	double-component line profiles}
        \label{f:mdotve}
\end{figure}

There are two main characteristics of the mass loss process, the
stellar mass loss rate and the circumstellar gas expansion velocity. 
The former is to a large extent determined by the conditions
at the transonic point, i.e., the density at the point where the gas
velocity goes from being sub- to supersonic, while the latter is
determined by the acceleration beyond this point.  Hence, these two
properties do not necessarily correlate with each other.
However, in a study of a dust-driven wind \citet{habietal94} found that
the relation $\dot{M}$$\propto$$v_{\rm e}$ should apply in the optically
thin limit.  Solving the same set of equations \citet{elitivez01} found
that the dependence becomes even stronger, $\dot{M}$$\propto$$v_{\rm
e}^{3}$, when the effect of
gravity is negligible. Therefore, a comparison between the two mass
loss characteristics may provide important results, which any mass
loss mechanism model must be able to explain.

In Fig.~\ref{f:mdotve} we present the mass
loss rates and gas expansion velocities for our sample.  There is
definitely a trend in the sense that the mass loss rate and gas
expansion velocity increase jointly.  A linear fit to the data results
in $\dot{M}$$\propto$$v_{\rm e}^{1.4}$ with a correlation coefficient
of 0.68.  For the C-star IRVs and SRVs the corresponding result is
$\dot{M}$$\propto$$v_{\rm e}^{2.0}$ with a correlation coefficient of
0.76.  Thus, the dependence is weaker for the M-stars, but this result
is hardly significant.  The spread in mass loss rate for a given
velocity is substantial, and larger than the estimated uncertainty in
the mass loss rate.  Results of similar nature have been found for other
samples of stars.  \citet{youn95} found a strong dependence,
$\dot{M}$$\propto$$v_{\rm e}^{3.4}$, for a sample of short-period
M-Miras, and \citet{knapetal98} found $\dot{M}$$\propto$$v_{\rm
e}^{2}$ for a sample containing a mixture of M- and C-stars. 
Differences in the slope may occur if different methods for estimating
mass loss rates have systematic trends, e.g., a systematic
underestimate at low mass loss rates and low expansion velocities, but also
the range of mass loss rates covered, the types of variables, etc.,
will have an effect.

Nevertheless, we can conclude that the mass loss mechanism(s) produces a
correlation between its two main characteristics which is in line with
theoretical predictions for an optically thin, dust-driven wind.  Our
result is also consistent with a rather weak dependence for low
luminosity sources where gravity effects cannot be ignored. The
considerable spread in mass loss rate for a given velocity
suggests that the outcome for an individual star is sensitive to the
conditions in the region where these properties are determined.

\subsection{Low gas expansion velocity sources}
\label{s:lowve}

The fraction of objects with low gas expansion velocities is high in
our sample.  There are 20 objects (i.e., 30\% of the whole sample) with
velocities lower than 5\,km\,s$^{-1}$.  The corresponding value among
the carbon star IRVs and SRVs is only 8\%.  There is no sample
selection bias for or against low velocity sources in any of the
samples, and since the detection rates are high for both samples, we
conclude that there is clearly a difference between the mass loss
properties of AGB-stars of M- and C-type in this respect.

\begin{figure}
       {\includegraphics[width=8.5cm]{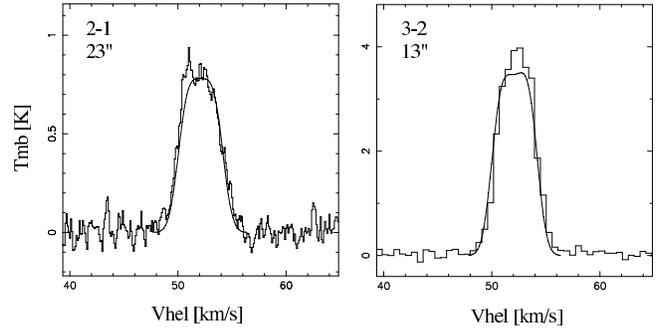}}
        \caption{CO($J$=2$\rightarrow$1) and CO($J$=3$\rightarrow$1)
        spectra (histograms) obtained at the SEST and the JCMT, respectively, and
        model line profiles (thin, solid lines) for the low gas expansion velocity
        source \object{L$^2$~Pup} (the beam size is given in each panel)}
        \label{f:l2pup}
\end{figure}

Of particular interest for further studies are the sources with gas
expansion velocities lower than 3\,km\,s$^{-1}$: \object{V584~Aql}, \object{T~Ari},
\object{BI~Car}, \object{RX~Lac}, and \object{L$^2$~Pup}.
Such a low velocity corresponds to the escape
velocity at a distance of 1.5$\times$10$^{15}$\,cm for a 1\,$M_{\odot}$
star, i.e., a distance corresponding to about 100 stellar radii,
considerably larger than the normally accepted acceleration zone,
about 20 stellar radii \citep{habietal94}.  In a detailed study of state
of the art mass loss models \citet{wintetal00b} concluded that for low
radiative acceleration efficiencies only low mass loss rates
($\lesssim$3$\times$10$^{-7}$\,M$_{\odot}$\,yr$^{-1}$) and low gas
expansion velocities ($\lesssim$5\,km\,s$^{-1}$) are produced (their
class B models).  In these models the gas expands at a relatively
constant and low velocity beyond a few stellar radii, and it finally
exceeds the escape velocity at large radii.  This can provide an
explanation for the low velocity sources, but also more complicated
geometries/kinematics may play an important role
(Sect.~\ref{s:doublecomp}).

\begin{figure*}    
\sidecaption
	{\includegraphics[width=12.0cm]{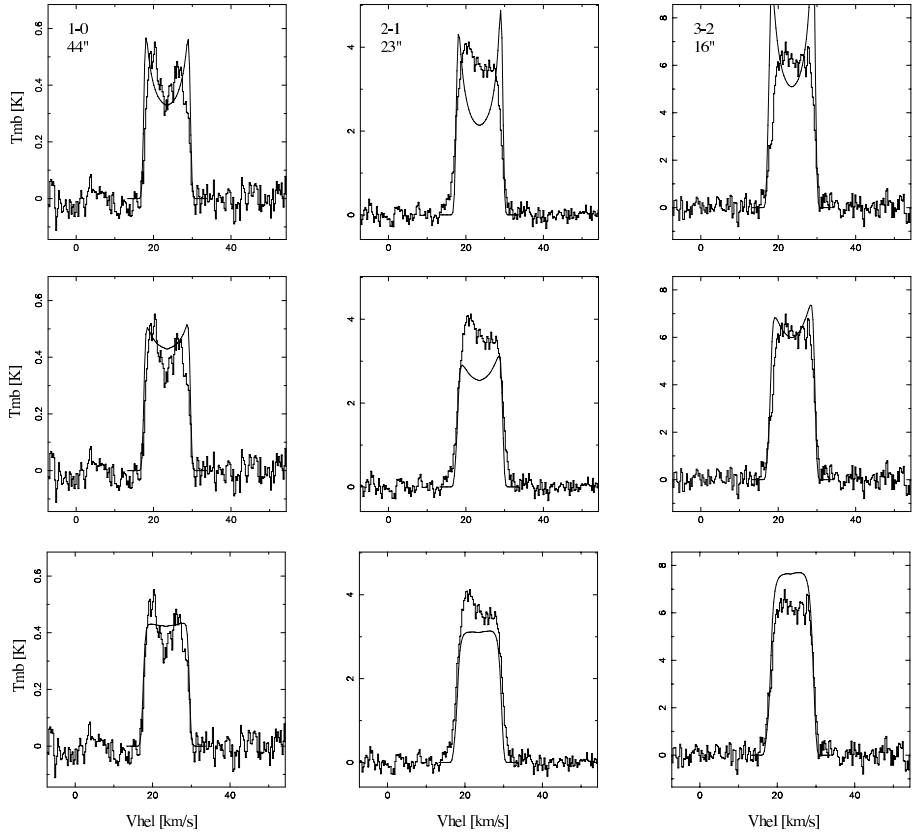}}
        \caption{Observed and modelled line profiles for \object{R~Dor}.  Upper panels
        show the results of the nominal model of \object{R~Dor} ($D$=46\,pc,
        $\dot{M}$=1.3$\times$10$^{-7}$\,M$_{\odot}$\,yr$^{-1}$, $h$=0.7).  The
        middle panels show the results for a distance three times larger
        ($D$=150\,pc, $\dot{M}$=5$\times$10$^{-7}$\,M$_{\odot}$\,yr$^{-1}$,
        $h$=1.5).  The lower panels show the results for a photodissociation
        radius three times smaller than that of the nominal model
        ($\dot{M}$=3$\times$10$^{-7}$\,M$_{\odot}$\,yr$^{-1}$, $h$=0.05)}
        \label{f:rdor}
\end{figure*}

\citet{juraetal02} presented a study of \object{L$^2$~Pup} where they used the
mid-IR cameras on the Keck telescope to partly resolve the dust
emission.  No clear geometrical structure is evident, but they derive
a dust mass loss rate of 10$^{-9}\,M_{\odot}$\,yr$^{-1}$, which
combined with our gas mass loss rate, results in a dust-to-gas mass
ratio as high as 0.05, i.e., about 15 times higher than our average
estimate from the $h$-parameter (Sect~\ref{s:h}).  This suggests that there are
problems with the dust and/or the gas mass loss rate estimates of this
star.  In Fig.~\ref{f:l2pup} we show the model line profiles
superimposed on our highest quality spectra of this object.  A very
good fit is obtained for a mass loss rate of
2.2$\times$10$^{-8}$\,M$_{\odot}$\,yr$^{-1}$, an expansion velocity of
2.1\,km\,s$^{-1}$, and a turbulent velocity width of
1.0\,km\,s$^{-1}$ (this is higher than the 0.5\,km\,s$^{-1}$ adopted in the
modelling of the whole sample).  However, this object is so extreme
that not too much faith should be put in the model results despite the
successful fit.

\subsection{The line profiles}
\label{s:lineprof}

The line profiles carry information which was not used to constrain
the derived mass loss rates.  In this section we briefly discuss how
well the model line profiles reproduce the observed ones.  In general,
the results are satisfactory. However, although there are both
too double-peaked and too rounded model profiles when compared with the
observed ones, there is a trend of too double-peaked (or too flat)
model line profiles, indeed observed double-peaked line profiles are
very rare.  The discrepancy is worst for the $J$=1$\rightarrow$0
line, where double-peaked model line profiles are common.  There are
two possibilities for such a discrepancy, either the angular size of
the emitting region is too large in the model or there is an effect
due to maser emission, which is not reproduced in nature.  The former
can be due to systematically too small distances to the sources or too
large photodissociation radii.  The latter is a possibility because
for these low mass loss rate objects radiative excitation plays an
important role and it tends to invert preferentially the lower
$J$-transitions.  However, among the objects with discrepancies there
are about as many without maser action as with maser action (in the
model).

\begin{table*}
    \centering
    \caption[]{Source parameters and model results for those objects
    with double component line profiles}
     \label{t:modresdoublecomp}
   \[
    \begin{tabular}{llrrclccclrc} 
    \hline
    \noalign{\smallskip}
Source & Var.  & P{\phantom{00}} & D & T$_{\rm bb}$ & comp.  & $\dot{M}$ & $v_{\rm e}$ & $r_{\rm p}$ & $h$ & $\chi^2_{\rm red}$ & N \\
 & type & [days] & [pc] & [K] & & [10$^{-7}$\,M$_{\odot}$\,yr$^{-1}$] & [km\,s$^{-1}$] & [10$^{16}$\,cm] & & &\\
\hline
EP Aqr   & SRb & 55  & 140 & 2200 & broad  & 5{\phantom{.0}} & 9.2  & 2.5 &  &      & 1 \\
         &     &     &     &      & narrow & 0.3             & 1.0  & 1.1 &  &      & 1 \\
RV Boo   & SRb & 137 & 280 & 2760 & broad  & 2.0             & 7.0  & 1.8 & 0.1  & 0.7  & 4 \\
         &     &     &     &      & narrow & 0.3             & 2.3  & 0.8 & 0.05 & 9.9  & 4 \\     
X Her    & SRb & 95  & 140 & 2490 & broad  & 1.5             & 6.5  & 1.5 & 0.2  & 4.6  & 3 \\
         &     &     &     &      & narrow & 0.4             & 2.2  & 1.0 & 0.03 & 4.5  & 3  \\
SV Psc   & SRb & 102 & 380 & 2450 & broad  & 3{\phantom{.0}} & 9.5  & 1.9 & 0.1 & 2.1  & 3 \\
         &     &     &     &      & narrow & 0.3             & 1.5  & 1.0 & 0.05 & 20.7  & 3 \\
\hline
     \end{tabular}
        \]
\end{table*}

The by far worst discrepancy is obtained for R~Dor, where all the
model line profiles are strongly double-peaked, Fig.~\ref{f:rdor}. 
All three transitions are inverted (the $J$=1$\rightarrow$0 line over
the whole radial range, but the $J$=2$\rightarrow$1 and
$J$=3$\rightarrow$1 lines only over a part of it), but the optical depths are
so low that substantial effects of maser action are not expected.  The
double-peakedness is rather due to the large angular extent of the
emission in the model, i.e., the emission is clearly spatially
resolved.  This can be solved by increasing the distance.  The
somewhat uncertain Hipparcos distance of R~Dor is 61\,pc, but a change
to this distance leads only to marginal improvements in the model
fitting.  We have to increase the distance by a factor of three to get
acceptable fits to the observed data ($D$=150\,pc for which the
best-fit results are
$\dot{M}$=5$\times$10$^{-7}$\,M$_{\odot}$\,yr$^{-1}$ and $h$=1.5; note
that the derived temperature structure depends on the distance to the
source due to the emission being spatially resolved).  Such a large
distance is not obviously compatible with the Hipparcos data. 
Alternatively, we can artificially lower the size of the CO envelope
by a factor of three compared to that given by the model of
\citet{mamoetal88} [for which the best fit results are
$\dot{M}$=3$\times$10$^{-7}$\,M$_{\odot}$\,yr$^{-1}$ and $h$=0.05; for
this mass loss rate the photodissociation radius is actually five
times larger according to the model of \citet{mamoetal88}].  Actual
tests of the predictions of the model of \citet{mamoetal88} have
mainly been done for high mass loss rate C-stars
\citep{schoolof00,schoolof01}.  The model has passed these tests, and
it is therefore questionable whether it gives results off by a factor
of five at lower mass loss rates.  A possibility exist that the
interstellar UV field is exceptionally strong and hence limits
severely the CO envelope of R~Dor. Alternatively, the mass loss is
highly variable, and the small outer radius of the CO envelope is a
consequence of a recent higher mass loss epoch. We must conclude that
presently the reason for the major discrepancy between the
observational data of R~Dor and our modelling results is not clear.  A
complicating factor is that the $J$=1$\rightarrow$0 line profile is
time variable (Nyman, priv.  comm.), and that it at times looks rather
peculiar [compare e.g. the spectrum shown in \citet{lindetal92c}].

Finally, we note here that \citet{olofetal93a} in their sample of
C-type SRVs and IRVs found five sources (about 10\%) with
remarkable CO radio line profiles, sharply double-peaked with only
very weak emission at the systemic velocity.  These were subsequently
shown to originate in large, detached shells \citep{olofetal96}, which
are geometrically very thin \citep{lindetal99,olofetal00}, presumably
an effect of highly episodic mass loss.  We have found no such source
in our sample of M-stars.  Hence, in this respect there must be a
difference in the mass loss properties between the two chemistries.

\subsection{Double-component line profiles}
\label{s:doublecomp}

\citet{knapetal98} and \citet{kersolof99} were the first to discuss more
thoroughly the small number of objects with line profiles which can be
clearly divided into two components, a narrow feature centred on a
broader plateau.  Such sources exist both among M- and C-type stars
\citep{knapetal98}.  The origin of such a line profile is still not
clear.  \citet{knapetal98} argued that it is an effect of episodic mass
loss with highly varying gas expansion velocities.  Alternatively,
it can be an effect of complicated geometries/kinematics. The first spatial
information was provided by \citet{kahajura96}.  A CO radio line map
towards the M-star X~Her suggested that the broad plateau is a bipolar
outflow, while the narrow feature was not spatially resolved. 
\citet{bergetal00} produced CO radio line interferometry maps of the
M-star RV~Boo.  In this case the brightness distributions suggest that
the broad plateau emission comes from a circumstellar disk with
Keplerian rotation.  \citet{kahaetal98} and \citet{jurakaha99}
interpret the narrow CO radio line features which they observe in a
few cases as originating in reservoirs of orbiting gas (these sources
do not have distinct double-component line profiles).  It is fair to
say that no consensus has been reached about these peculiar
circumstellar emissions.

In our sample we have four sources of this type, \object{EP~Aqr},
\object{RV~Boo}, \object{X~Her}, and \object{SV~Psc}, all of them SRVs.  We have simply
decomposed the emission into two components for each source, assuming
that the emissions are additive.  Mass loss rates and gas expansion
velocities were determined in the same way as for the rest of our
objects.  This is probably a highly questionable approach for both
components.  The results, as well as some source information, are
given in Table~\ref{t:modresdoublecomp}.  \citet{knapetal98}
derived mass loss rates for \object{EP~Aqr} and \object{X~Her} which
are within a factor of two of our estimates (both for the narrow and the broad
components).

Not unexpectedly the mass
loss rates are higher for the broader component by, on average, an
order of magnitude.  The fits to the narrow components result in very
low gas expansion velocities.  Indeed, so low, e.g., 1.0\,km\,s$^{-1}$
in the case of \object{EP~Aqr}, that an interpretation in the form of a
spherical outflow is put to question.  On the other hand, the relation
between mass loss rate and gas expansion velocity is the same for
these objects as for the rest of the sources, Fig.~\ref{f:mdotve}. 
The narrow component gas appears cooler than the broad emission gas in
those three cases where an $h$ can be estimated.  This may be an
accidental result, but it can also provide a clue to the
interpretation.

\section{Conclusions}

We have determined mass loss rates and gas expansion velocities for a
sample of 69 M-type IRVs (22 objects) and SRVs (47 objects) on the
AGB using a radiative transfer code to model their circumstellar CO radio line
emission.  We believe that this sample is representative for the mass
losing stars of this type.  The uncertainties in the estimated mass
loss rates are rather low within the adopted stellar/circumstellar
model, typically less than $\pm$50\%.  However, a sensitivity analysis
shows that for these low mass loss rate stars there is a considerably
uncertainty due to the stellar luminosity, the size of the CO
envelope, the CO abundance, and as usual the distance to the source. 
We find that the mass loss rates determined by the detailed radiative
transfer analysis differ by almost an order of magnitude from those
obtained by published mass loss rate formulae.

The (molecular hydrogen) mass loss rate distribution has a median value of
2.0$\times$10$^{-7}$\,M$_{\odot}$\,yr$^{-1}$, and a minimum of
2.0$\times$10$^{-8}$\,M$_{\odot}$\,yr$^{-1}$ and a maximum of
8$\times$10$^{-7}$\,M$_{\odot}$\,yr$^{-1}$.  M-type IRVs and SRVs with
a mass loss rate in excess of
5$\times$10$^{-7}$\,M$_{\odot}$\,yr$^{-1}$ must be very rare, and in
this respect the regularity and amplitude of the pulsation plays an
important role. We also find that among these mass losing stars the number of
sources with mass loss rates below a few
10$^{-8}$\,M$_{\odot}$\,yr$^{-1}$ must be small.

We find no significant difference between the IRVs and the SRVs in
terms of their mass loss characteristics.  Among the SRVs the mass
loss rate shows no dependence on the period.  Thus, for these
non-regular, low-amplitude pulsators it appears that the pulsational
pattern plays no role for the mass loss efficiency.

We have determined temperatures for our sample stars by fitting
blackbody curves to their spectral energy distributions.  These
blackbody temperatures have been shown to correlate reasonably well
with the stellar effective temperatures.  The mass loss rates of our
stars show no correlation at all with these stellar blackbody temperatures.

The gas expansion velocity distribution has a median of
7.0\,km\,s$^{-1}$, and a minimum of 2.2\,km\,s$^{-1}$ and a maximum of
14.4\,km\,s$^{-1}$.  No doubt, these objects sample the low gas expansion
velocity end of AGB winds.  The fraction of objects with low gas
expansion velocities is high, about 30\% have velocities lower than
5\,km\,s$^{-1}$.  There are four objects with gas expansion velocities
lower than 3\,km\,s$^{-1}$: \object{V584~Aql}, \object{T~Ari},
\object{BI~Car}, \object{RX~Lac}, and \object{L$^2$~Pup}. 
These objects certainly deserve further study.

We find that the mass loss rate and the gas expansion velocity
correlate well, $\dot{M}$$\propto$$v_{\rm e}^{1.4}$, even though for
a given velocity (which is well determined) the mass loss rate may take
on a value within a range of a factor of five (the uncertainty in the
mass loss rate estimate is lower than this within the adopted
circumstellar model). The result is in line with theoretical
predictions for an optically thin, dust-driven wind.

A more detailed test of the CO modelling is provided by the shape of
the line profiles.  In general, the fits are acceptable, but there is
a trend that the model profiles, in particular the $J$=1$\rightarrow$0
ones, are more flat-topped, or even weakly double-peaked, than the
observed ones.  An exceptional case is \object{R~Dor}, where the high-quality,
observed line profiles are essentially flat-topped, while the model
ones are sharply double-peaked.  Acceptable fits are obtained  by increasing the
distance to the star or by artificially decreasing the size of the CO
envelope.

The sample contains four sources with distinctly double-component CO
line profiles: \object{EP~Aqr}, \object{RV~Boo}, \object{X~Her}, and
\object{SV~Psc} (all SRVs).  We have modelled the two components
separately for each star and derive mass loss rates and gas expansion
velocities using the same circumstellar model as for the rest of the
sample.  The resulting mass loss rates and gas expansion velocities
show the same positive correlation as that of the other objects.  At
present, the exact nature(s) of these objects is unknown.

We have compared the results of this M-star sample with a similar
C-star sample.  The mass loss rate distributions are comparable,
suggesting no dependence on chemistry for these types of objects. 
Likewise, the mass loss rates of the C-stars show no correlation with
stellar temperature or period. The gas
expansion velocity distributions though are clearly different.  The fraction
of low velocity sources is much higher in the M-star sample.  In both
cases there is a correlation between mass loss rate and gas expansion
velocity, although the detailed relations are different.  Our crude
estimates of the dust properties, through the gas-grain collision
heating term, indicate that the two samples have similar gas-to-dust
ratios and that these differ significantly from that of high mass loss
rate C-stars.  This also means that the gas-CSEs due to low mass loss
rates are cooler than expected from a simple extrapolation of the
results for \object{IRC+10216}.  Finally, we find no example of the
sharply double-peaked CO line profile, which is evidence of a large,
detached CO-shell, among the M-stars.  About 10\% of the C-stars show
this phenomenon.

\begin{acknowledgements}
We are grateful to the referee, M.~Marengo, for a careful reading of
the manuscript and for constructive comments.  We also thank L.-{\AA}.  Nyman
for giving us the R~Dor CO($J$=$1\rightarrow$0) spectrum obtained at
the SEST. Financial support from the Swedish Science Research Council
is gratefully acknowledged.  FK was supported by APART (Austrian
Programme for Advanced Research and Technology) from the Austrian
Academy of Sciences and by the Austrian Science Fund Project
P14365-PHY. FLS was supported by the Netherlands Organization for
Scientific Research (NWO) grant 614.041.004.
\end{acknowledgements}

\bibliographystyle{aa}

\end{document}